\documentclass[11pt]{article}
\usepackage{moriond,epsfig}

\usepackage[brazil]{babel}
\usepackage[latin1]{inputenc}

\def\be{\begin{equation}}
\def\ee{\end{equation}}
\def\bea{\begin{eqnarray}}
\def\eea{\end{eqnarray}}

\begin{document}
\vspace*{4cm}
\title{HOLOGRAPHIC MODELS OF ELECTROWEAK SYMMETRY BREAKING
\footnote{Talk presented at the 43rd Rencontres de Moriond, La Thuile,
March 1-8 2008.}}

\author{ GUSTAVO BURDMAN }

\address{Physics Institute, University of São Paulo, 
R. do Matão 187, Travessa R,\\
Cidade Universitaria, São Paulo, Brazil}

\maketitle\abstracts{
We review the status of models of electroweak symmetry breaking in a slice 
of anti--de Sitter space. These  models can be thought of as dual to strongly 
interacting theories of the electroweak scale. 
After an introduction to some generic issues in bulk theories in AdS$_5$,
we concentrate on the model-building of the Higgs
sector. }

\section{The Hierarchy Problem and Strong Dynamics}
As the Large Hadron Collider (LHC) gets close to start taking data, we 
continue to ponder what new physics could appear at the TeV scale.
In the standard model (SM) the electroweak symmetry is broken by a scalar doublet. 
This implies the existence of an elementary Higgs boson that must be relatively 
light ($<1~$TeV) to unitarize electroweak scattering amplitudes, even lighter to satisfy 
electroweak precision constraints. However, its mass, and with it the electroweak scale, 
is unstable under radiative  corrections. In order to keep it below the TeV scale and close 
to $v\simeq 246~$GeV, the bare mass parameter (presumably controlled by ultra-violet physics)
must be finely adjusted to cancel against quadratically divergent 
loop corrections driven by the SM states. The need for this cancellation is highly unnatural 
and is called the 
hierarchy problem. In order for naturalness to be restored, new physics must cancel the quadratic 
divergences at a scale not far above the TeV scale. 

The solution of the hierarchy problem is likely to shed light on the 
origin of electroweak symmetry breaking  (EWSB). We can classify scenarios of new physics
beyond the SM by how they solve the hierarchy problem. For instance, in 
supersymmetric theories at the weak scale~\cite{smartin}, the quadratic divergences in the Higgs mass squared
are canceled by the contributions of super-partners of the SM particles.
Soft SUSY breaking allows for enough contributions to $m_h^2$ to trigger EWSB radiatively. 

An alternative to solve the hierarchy problem is the possibility that some sort of new strong dynamics
is present at or just above the TeV scale~\cite{strongd}. 
For instance, in Technicolor theories~\cite{tc1,tc2} the new strong interaction becomes strong enough at the 
TeV scale to trigger the condensation of Techni-fermions. If some of these are $SU(2)_L$ doublets, this 
triggers EWSB. There is no Higgs in this type of QCD-like scenario. The hierarchy problem is solved
by dimensional transmutation, i.e. just as in the strong interactions the coupling becomes 
strong at low energies triggering EWSB naturally. The TeV scale is encoded in the running of this 
new strong coupling in the same way that the typical scale of hadronic physics ($\sim 1~$GeV) is 
encoded in the running of $\alpha_s$.

The trouble starts when 
one tries to construct the operators responsible for fermion masses. For this purpose one needs
to extend the gauge group in Extended Technicolor (ETC) models. If at some 
energy scale $\Lambda_{\rm ETC}$ the ETC gauge bosons acquire masses, they generate 
four-fermion operators involving both fermions and Techni-fermions:
\be
\frac{g_{\rm ETC}^2}{M_{\rm ETC}^2}\,\bar f_L f_R \bar T_R T_L 
\ee
At the lower TC scale $\Lambda_{\rm TC}$, the formation of the Techni-condensate 
$\langle \bar T_L T_R\rangle\sim \Lambda_{\rm TC}^3$ results in a fermion mass of the order of 
\be
m_f \simeq \frac{g_{\rm ETC}^2}{M_{\rm ETC}^2} \, \Lambda_{\rm TC}^3
\ee
Clearly, if one wants to explain the fermion mass hierarchy one needs either several 
very different ETC scales, or some non-standard type of dynamics, most probably both.
Particularly troublesome for ETC models are heavier masses, say above the charm 
mass. To obtain the top mass the ETC mechanism fails given than the ETC scale should be right on top of
the TC scale. 

Several complications of the ETC/TC idea allow for the fermion mass hierarchy 
(tumbling~\cite{tumble}, walking TC~\cite{walk}) and even for the top quark mass (top-color assisted TC~\cite{toptc}). 
In any case, it is clear that the dynamics associated with TC/ETC models must be quite 
different from that of a simple scaled up QCD-type theory. In addition to their problems 
with the generation of fermion masses, scaled up QCD TC models tend to predict a rather large 
S parameter 
\be
S \simeq \frac{N}{6\pi}
\ee  
where $N = N_{\rm tc} N_{\rm D}$ is the product of the number of
Techni-colors and the number of weak doublets.

Despite all these problems, the idea that a new strong interaction at or just above the TeV scale
is responsible for EWSB (and solves the hierarchy problem) remains attractive.
Perhaps the new  strong interaction is quite different from QCD and we do not have yet neither the 
theoretical tools  nor the experimental guidance needed to build it.

\section{Building Strongly Coupled Theories in AdS$_5$}

Fortunately, there is a way to build a large array of strongly coupled theories 
of EWSB by using the AdS/CFT correspondence~\cite{adscft}.  
The conjecture relates a Type~IIB string theory on AdS$_5\times S^5$
with a four-dimensional  ${\cal N} =4$ SU(N) gauge theory, which is a
conformal field theory. By extension, we can assume that at low
energies we can describe the string theory  by a higher dimensional field theory.
As long as the AdS radius $R_{\rm AdS}$ is much larger than the string
scale $\ell_s$ the description in the higher-dimensional theory is
weakly coupled. On the other hand, this leads to ${\tilde g} N\gg 1$,
where $\tilde g$ is the gauge coupling of the Yang-Mills theory. 
Then, the description of a weakly coupled theory in AdS$_5$
corresponds to a strongly coupled four-dimensional theory. We can
think of the large $N$ limit of the 4D gauge theory in terms of planar
diagrams, which in turn are reminiscent of a loop expansion in the
topology of the world-sheet in string theory. 

In the original AdS/CFT correspondence, the boundary of AdS space is
set at infinity, and is just a Minkowski 4D boundary. The 
corresponding 4D theory is exactly conformal, and therefore is not
suitable for building models of EWSB. For our purposes, we want to 
consider an ultra-violet (UV) cutoff of the 4D theory, corresponding
to a UV boundary at a finite coordinate in the extra dimension. Also, 
in order to obtain the description of an interesting strongly coupled
theory, we require that the 4D gauge theory leads to non-trivial
dynamics which triggers EWSB. This infra-red (IR) physics can be mimicked
in the 5D theory by the appearance of an IR boundary. Thus, if we put
the UV boundary at the origin of the extra dimension, $y=0$,  the
statement of the AdS/CFT reads
\bea
&&\int [D\phi_0] e^{i\,S_{\rm UV}[\phi_0]}\,\int [D\phi_{\rm CFT}]
e^{i\,S_{\rm CFT}[\phi_{\rm CFT}] + i\int d^4x \phi_0 {\cal O}} \nonumber\\
&&= 
\int [D\phi_0] e^{i\,S_{\rm UV}[\phi_0]}\,\int_{\phi_0} [D\phi] e^{i
S_{\rm bulk}[\phi]}
\label{corresp}
\eea
where $\cal O$ is an operator in the strongly coupled 4D theory,
and $\phi_0\equiv \phi(x,y=0)$ is a UV boundary field which acts as a
source for the 4D operator $\cal O$. The correspondence in
Eq.~(\ref{corresp}) states the the 4D action is equivalent to the 
effective action for the source field $\phi_0$ on the UV boundary,
which is obtained by integrating over the bulk degrees of freedom of
$\phi(x,y)$: 
\be
e^{i\,S_{\rm eff.}[\phi_0]} = \int_{\phi_0} [D\phi] e^{i
S_{\rm bulk}[\phi]}
\ee
Then, n-point functions of the 4D theory can be obtained by using
$S_{\rm eff.}[\phi_0]$ as a generating functional. It is in this sense
that the bulk degrees of freedom are determining the dynamics of the
4D theory.

We then see that for us to build models of strongly coupled 
theories, for instance to explain EWSB, we must specify a weakly
coupled 5D theory in AdS$_5$. The type of strongly coupled theories
that we can build this way is not completely general. For instance, it requires 
that it be ``large N'', which should in principle translate in the
presence of narrow resonances.  
In what follows we consider the steps to build such theories.

\subsection{Solving the Hierarchy Problem in a slice of AdS$_5$}
The starting point to build theories of EWSB using holography is 
the Randall-Sundrum setup as a 
solution of the hierarchy problem~\cite{rs1}.  
We consider an extra dimension compactified on an orbifold, i.e. $S_1/Z_2$, with 
the metric
\be
ds^2 = e^{-2\,k\,y} \eta^{\mu\nu} dx_\mu dx_\nu - dy^2~,
\label{metric}
\ee
where $k \sim M_P$ is the AdS curvature. The orbifold compactification $S_1/Z_2$ results in 
a slice of AdS in the interval $[0,\pi R]$, with $R$ the compactification radius.
This metric is a solution of Einstein's equations if we fine-tune the bulk cosmological constant
to cancel the brane tensions. 
This choice of metric means that the graviton's wave-function is exponentially 
suppressed away from the origin. 
In general, this metric 
exponentially suppresses all energy scales away from 
the origin. Then, if the Higgs field is localized a distance $L=\pi R$ from the origin, 
\be  
S_{H} = \int\,d^4x\int_{0}^{\pi R} dy\,\sqrt{g}\,{\delta(y-\pi R)} \left[
g^{\mu\nu}\partial_\mu H^\dagger \partial_\nu H - \lambda\left(|H|^2 - {v_0^2}\right)^2
\right]
\ee
where $\lambda$ is the Higgs self-coupling and $v_0$ is its vacuum expectation value (VEV). 
The latter must satisfy $v_0 \sim k$ for the theory to be technically natural. 
Taking into account the exponential factors in $g^{\mu\nu}$ and $\sqrt{g}$, 
the renormalization of the Higgs field required to render its kinetic term 
canonical results in an effective four-dimensional VEV given by 
\be
v = e^{-k\pi R} \,v_0
\ee
Thus, in order for the weak  scale $v\simeq 246~$GeV 
to arise at the fixed point in $y=\pi R$, we need $k R\sim (10-12)$.
Then, if the Higgs is for some reason localized at or nearly at this location, 
this setup constitutes a solution to the hierarchy problem. 

In the original Randall-Sundrum (RS) proposal only gravity propagates in the extra dimension.
However, this presents several problems, mostly associated with the fact that is not 
possible to sufficiently suppress higher-dimensional operators. For instance, operators mediating 
flavor violation might only be suppressed by the TeV scale, $k\,e^{-k\pi R}$. 
Grand Unified Theories (GUTs) might not be viable since we cannot effectively suppress proton decay. 
Many of these problems are solved when fermions and gauge bosons are allowed in the AdS$_5$ 
bulk. In fact, in order to solve the hierarchy problem, the only field that must remain localized 
near the $\pi R$ or TeV brane is the Higgs. In addition, allowing the standard model fields in the bulk
opens up a large number of model building possibilities including viable GUTs~\cite{alexgut} 
and the modeling of 
the origin of flavor~\cite{gp,gn} just to mention two very prominent cases. But most importantly, 
it gives us a tool
to build models of EWSB that address its dynamical origin.  
In building bulk models of EWSB we must dynamically explain why is the Higgs localized near the 
TeV brane. Building such models is like constructing strongly coupled theories of EWSB but from a 
different perspective. 

\subsection{Bulk RS Theories and the Origin of Flavor}

Writing a theory in the bulk require several ingredients. 
First, we must decide what the gauge symmetry should be. 
It turns out that the SM electroweak symmetry, $SU(2)_L\times U(1)_Y$, is not suitable 
for this since it results in a large value of the parameter $T$. 
The reason is that there is large explicit isospin violation in the bulk: in addition 
to the SM-sanctioned isospin violation proportional to $g'/g = \tan\theta_W$, the bulk 
adds the isospin violation of the Kaluza-Klein (KK) modes of the SM gauge fields. 
As a consequence, it is necessary to implement a gauge symmetry that would exhibit  
isospin symmetry in the bulk. A minimal extension of the SM with this feature is 
$SU(2)_L\times SU(2)_R\times U(1)_X$. This bulk symmetry may be broken by boundary conditions
to the SM gauge group or directly to $U(1)_{\rm EM}$.
Additional discrete symmetries may be imposed to protect the $Z\to \bar b b$ coupling from 
large deviations~\cite{o4}. 

One important consequence of writing a bulk theory, is that the KK states
would start from masses of the order of $M_{\rm KK} \sim O(1)~$TeV, independently of whether 
they correspond to fermions, gauge bosons or even scalars. This also means that the wave-function
of the KK excitations in the extra dimension also peaks at $y=\pi R$, i.e. at the TeV brane.
This is related to the fact that position in the bulk can be thought of  as an energy scale, and is 
a generic feature independent of the model considered.

Another important issue is the localization of zero modes in the bulk, i.e. their effective
5D wave-function. The strength of couplings between zero-modes and KK modes and, since the 
Higgs is TeV-localized, of the zero-mode fermion Yukawas, are determined by this. 
Fermions  propagating in the bulk can have a mass term (as long as we assume is an odd mass term). 
The natural order of magnitude of this fermion bulk mass is $k$, the only dimensionfull bulk parameter. 
We can then write the bulk fermion mass as
\be
M_f = c_f \, k
\ee
where $c_f\simeq O(1)$. Expanding the 5D fermion $\Psi(x,y)$ in KK modes and solving the 
equation of motion for the zero mode we see that its bulk wave-function, for instance for a 
left-handed zero mode, behaves like~\cite{gp}
\be
F_{\rm ZM}^L(y) \sim e^{(\frac{1}{2} - { c_L})\,k y}
\ee   
Then, if $c_L>1/2$ the zero-mode fermion is localized towards the Planck brane, whereas if 
$c_L<1/2$ it would be localized near the TeV brane. For a right-handed zero mode, localization 
near the Planck bran occurs if $c_R<-1/2$, and near the TeV brane if $c_R>-1/2$. 
If a zero mode fermion has a large wave-function near the TeV brane it has also a large overlap 
with the Higgs, which has to be localized there. Thus, heavier fermions 
must have a wave-function towards the TeV brane, whereas light fermions must be essentially 
Plank-brane localized in order to explain their small Yukawa couplings.
We see then that fermion localization in the AdS$_5$ bulk provides a potential 
explanation for the hierarchy of fermion masses in the SM. 
\begin{figure}
\begin{center}
\epsfig{file=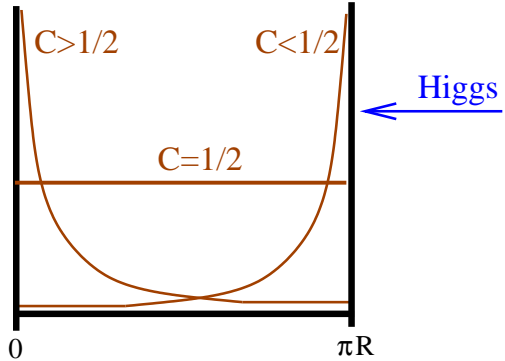,width=2.60in,angle=0}
\end{center}
\label{ferloc}
\caption{Localization of left-handed fermion zero-modes by variations
of $O(1)$ in the bulk mass parameter. }
\end{figure}
This emerging picture of flavor requires that the top quark be 
highly localized towards the TeV brane. Furthermore, the left-handed third-generation 
doublet contains the left-handed b quark, which then also has to have a substantial 
TeV localization. The rest of the zero-mode fermions must be localized mostly near the 
Plank brane. Since the zero-mode gauge bosons are flat in the extra dimension, this 
does not affect the universality of the zero-mode gauge couplings. However, 
the KK modes of gauge bosons are TeV-brane localized, and then couple stronger to the 
TeV-brane localized fermions, i.e. the third generation. 
The couplings of light fermions to KK gauge bosons are nearly universal, making 
low energy flavor phenomenology viable, even in the presence of tree-level flavor violation.
Also since light fermions are localized near the Plank brane, the scale suppressing higher dimensional 
operators responsible for proton decay is again $M_P$. 

Then any sign of the characteristic  tree-level flavor violation would have to appear 
in the interactions of the third generation quarks with the KK gauge bosons. 
Although these would have potentially important effects in flavor physics~\cite{flavors1,flavors2}, 
a direct
observation of the KK gluon decay into a single top and a jet, coming most likely from 
$G^{(1)} \to t \bar c$, would be an unambiguous signal of this theory of flavor~\cite{abe}.

Finally, a word about electroweak precision constraints. Since the bulk gauge theory has 
an isospin symmetry built in, we need not worry about the theory
generating a large $T$ parameter. 
However, this class of models all share the same problem with the $S$ parameter. They have an 
$S$ parameter which is approximately~\cite{adms,bn,csakis} 
\be
S_{\rm tree} \simeq 12\pi \frac{v^2}{M_{\rm KK}^2},
\label{stree}
\ee
which results in a bound of about $M_{\rm KK}>2.5~$TeV. 
This is a feature we have to live with in most bulk RS constructions. 
We can interpret this in the dual 4D picture, as the fact that in the 4D theory we have a 
large $N$, which enter in $S_{\rm tree}\sim \frac{N}{\pi}$, where $N$ is typically the size of the 
4D gauge group. This can in principle be made considerably smaller if the 
light fermions are allowed out of the Plank brane region and have almost flat profiles in the extra 
dimensions. Essentially this decouples them from the KK modes and avoids $S_{\rm tree}$~\cite{adms}.
But this picture would lack a solution to the fermion mass hierarchy.

\section{Electroweak Symmetry Breaking from AdS$_5$} 

We have arrived at a general picture of these kind of models in AdS$_5$: 
\begin{itemize}
\item The Higgs is TeV-brane localized in order to solve the gauge hierarchy problem
\item Fermion localization explains the fermion mass hierarchy: light fermions 
are Plank-brane localized resulting in small Yukawa couplings. Heavier fermions are TeV-brane 
localized ($t_R$, $t_L$ and $b_L$). 
\item The bulk gauge symmetry must be enlarged to protect isospin, to be at least 
$SU(2)_L\times SU(2)_R\times U(1)_X$.
\end{itemize}
\begin{figure}
\begin{center}
\epsfig{file=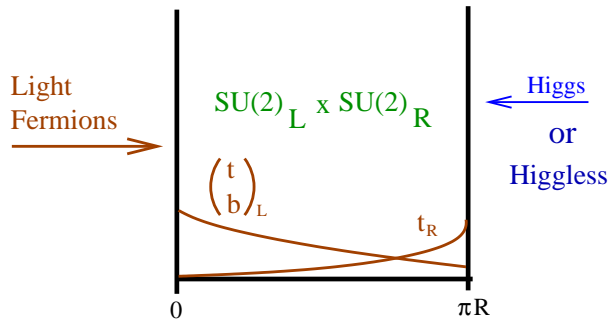,width=3.20in,angle=0}
\end{center}
\label{newmods}
\caption{Cartoon representation of bulk RS models.}
\end{figure}
This is already quite a rich structure with a very interesting 
phenomenology and lots of model building possibilities beyond EWSB and fermion masses. 
However, if these models are dual to a strongly coupled theory in four dimensions, as the presence of 
the resonances (i.e. the KK modes) appears to suggest, then we must be a bit disappointed with the 
Higgs sector. What keeps the Higgs 
localized to the TeV brane ?  Do we have to have a Higgs at all ? 
In what follows we briefly discuss three possibilities for the Higgs sector in these 
theories.

\subsection{Higgsless EWSB in AdS$_5$}
In the absence of a Higgs, the bulk gauge symmetry must be broken directly to $U(1)_{\rm EM}$
by boundary conditions (BC) for the bulk gauge fields at the fixed point of the extra dimension. 
On the Plank brane the BC are such that the bulk gauge symmetry breaks as~\cite{cccust} 
\be
SU(2)_R\times U(1)_X \longrightarrow U(1)_Y , ~~~{\rm at ~}y=0
\ee
On the other hand, on the TeV brane
\be
SU(2)_L\times SU(2)_R \longrightarrow SU(2)_V , ~~~{\rm at ~}y=\pi R
\ee
in such a way that it it preserves custodial symmetry. 
Then the gauge symmetry in the bulk has a gauged version of the SM custodial symmetry. 
This remains as a remnant global symmetry, resulting in the correct  masses for the $W$ and the $Z$.

More problematic in these models is how to give fermions their masses. 
Zero-mode fermions can obtain isospin conserving masses through a TeV-brane localized mass
term. Since $SU(2)_V$ is not broken there, isospin splitting is not generated by these terms. 
In order to achieve the freedom to have the correct mass spectrum one must introduce 
the following bulk spectrum (schematically and only for the third generation case):
\bea
\Psi_L &=& (t_L,b_L) ~~~({\bf 2,1})_{1/6} \nonumber\\
\Psi_R &=& (t_R,b'_R) ~~({\bf 1,2})_{1/6} \\
\Psi'_R &=& (t'_R,b_R) ~~({\bf 1,2})_{1/6} \nonumber
\eea
Each of these bulk fermions contains both left and right handed components.
We can choose the BCs so that the zero-modes of $\Psi_L$ correspond to the third generation 
left-handed quark doublet, the zero mode of $\Psi_R$ is the right-handed top $t_R$, and 
the one corresponding to $\Psi'_R$ is $b_R$. Mass terms localized in the TeV brane 
gives rise to fermion masses. It is still true that the larger the zero-mode wave-function is at the 
TeV brane, the larger its mass would be. But the top mass cannot be so easily adjusted. 
The reason is that there is a tension between the TeV localization of the top, which if 
too extreme produces noticeable deviation in the $Z \bar b_L b_L$ coupling, and the size of the 
isospin conserving TeV localized mass. The latter cannot be too large or it would induce a 
large mixing between $b'_R$ and the b-quark's zero mode through the mass term responsible for the 
top quark mass. This would again result in a deviation of $Z \bar b_L b_L$.
A way to circumvent this problem is to extend the custodial symmetry by a discrete symmetry~\cite{o4}, 
$P_{\rm LR}$ that relates the two $SU(2)$'s. In order to protect the $b_L$ coupling, it must be included 
in a bi-doublet of $SU(2)_L\times SU(2)_R$. Then the right-handed fermions could be in singlets or 
triplets of the $SU(2)$'s. For instance, if $t_R \sim ({\bf 1,1})_{2/3}$, then it does not contain
any fermion that could mix with the left-handed b. On the other hand, the field resulting in the 
right-handed zero mode would have to be a full $O(4)$ triplet $\Psi_R \sim ({\bf 3,1})_{2/3}
\oplus ({\bf 1,3})_{2/3}$ resulting in a distinct spectrum of KK fermions~\cite{lightcust}. 

But the most important signal of this scenario stems from the absence of the Higgs as a unitarizing
field in $VV$ scattering~\cite{wwcollider}, with $V = W^\pm, Z$. The unitarization of these amplitudes, unlike in 
other strongly coupled theories such as (4D) Techni-color, is the result of the presence of the 
narrow resonances that are the KK modes of the gauge bosons. The constraint of unitarization imposes 
sum rules on the couplings. For instance, 
\bea
{ g_{WWWW}}&=& {g_{WWZ}^2} + { g_{WW\gamma}^2} + 
\sum_n ({g_{WWV^{(n)}}})^2 \nonumber\\
 &=& \frac{3}{4 M_W^2}\,\left[ { g_{WWZ}}^2 M_Z^2 
 + \sum_n ({g_{WWV^{(n)}}})^2 M_n^2 \right]\nonumber
\eea
The requirement of unitarity of gauge boson scattering amplitudes means that the KK modes of gauge 
bosons cannot be too heavy. For example, if one wants to preserve perturbative unitarity, they must be 
below the TeV scale. Higgsless models in AdS$_5$ are then characterized by relatively low mass 
KK excitations.

\subsection{Gauge-Higgs Unification}

A remarkable mechanism to obtain a Higgs field naturally localized near the TeV, 
naturally light and suitable for EWSB is that in which the Higgs comes from 
a gauge field in 5D. In general, a 5D gauge field $A_M(x,y)$, $M=0,1,2,3,5$, can be decomposed 
in a vector $A_\mu(x,y)$ and a scalar component $A_5(x,y)$. 
If we want to extract the Higgs $SU(2)_L$ doublet from a gauge field in 5D, then the 
gauge symmetry in 5D has to be enlarged beyond the SM gauge symmetry. In order to illustrate
how this works let us take a simple example, an $SU(3)$ bulk gauge theory~\cite{gh1}. 
We can use BCs to break this gauge symmetry as $SU(3)\to SU(2)_L\times U(1)_Y$.
By choosing the BCs appropriately, the gauge fields are
\bea
{t^aA_\mu^a}:\: \left( \begin{array}{cc|c}
    (+,+) & (+,+) & (-,-) \\ 
    (+,+) & (+,+) & (-,-) \\ \hline
    (-,-) & (-,-) & (+,+) 
\end{array}\right)
\nonumber\\
\nonumber\\
  {t^aA_5^a}:\: \left( \begin{array}{cc|c}
    (-,-) & (-,-) & (+,+) \\ 
    (-,-) & (-,-) & (+,+) \\ \hline
    (+,+) & (+,+) & (-,-) 
  \end{array} \right)
\nonumber
\eea
where $a=1,2,3$ is the $SU(3)$ adjoint index and $t^a$ are the $SU(3)$ generators, and we use the fact 
that the BCs for the $A_5^a(x,y)$ are always opposite from those for $A_\mu^a(x,y)$. 
The signs correspond to the BCs in the Plank and TeV branes respectively. 
Thus, we see that the spectrum of zero-mode gauge bosons corresponds to $SU(2)\times U(1)$ as the 
gauge symmetry. The symmetry has been reduced or broken by this choice of BCs. 
Furthermore, we see that the $A_5^a(x,y)$'s corresponding to the ``broken'' generators, i.e. the generators
for which $A_\mu^a(x,y)$ does not have a zero mode,  have zero modes (i.e. $(+,+)$ BCs). 
In fact, these constitute four real degrees of freedom that can be seen to be a doublet of 
$SU(2)$ and its adjoint. Then, we can identify this $SU(2)$ doublet with the Higgs. 
In the case of an AdS$_5$ metric, if we impose the unitary gauge, this results in 
\be
\partial_y (k e^{-ky}\,A_5(x,y)) = 0, ~
\ee
which results in a scalar doublet with a profile localized towards the TeV brane. 
Thus, we achieved Higgs TeV-brane localization and extracted the Higgs from a 
gauge field in the bulk. These models are clearly related to Little Higgs theories, where the 
Higgs is a (pseudo-)Nambu-Goldstone boson (pNGB): the Higgs is associated to the broken generators
of a symmetry, which from the 4D interpretation would be a global one. It cannot have a potential since
shift symmetry (the remnant gauge symmetry for the $A_5^a$ zero-modes)  forbids it. 
Thus, these models should be dual to 4D theories of a composite Higgs, where the Higgs is 
a pNGB. 

This simple $SU(3)$ model of Gauge-Higgs unification has a lot of the features that we want. 
However, it does not have custodial symmetry in the bulk. 
The way to cure this is to simply enlarge the bulk gauge symmetry. The minimal realistic model 
of Gauge-Higgs unification in AdS$_5$ requires that we start with $SO(5)\times U(1)_X$ broken
down to  $SO(4)\times U(1)_X$ on the TeV brane, whereas it is reduced to $SU(2)_L\times U(1)_Y$
on the Plank brane~\cite{gh2}. Just as in the previous case, in the unitary gauge only the $A_5^a(x,y)$'s 
associated with the broken generators, i.e. transforming in the coset space $SO(5)/SO(4)$, 
have zero modes. These are arranged in a ${\bf 4}$ of $SO(4)$, or a bi-doublet of
$SU(2)_L\times SU(2)_R$. 

Fermion masses are not very problematic and can be obtained by localization just as in the generic models
with a TeV-brane localized Higgs. 
Regarding electroweak precision constraints, these models can evade
them more efficiently.
In particular, the $S$ parameter can be made in agreement with experiment for gauge 
KK  masses above 2~TeV. On the other hand, KK fermions could be quite a bit lighter, especially once 
the additional discrete custodial symmetry is introduced in order to keep the $Z\to \bar b_L b_L$ in check.
For instance, possible embeddings  are~\cite{ghcust} 
\be
{\bf 5_{2/3}} = ({\bf 2},{\bf 2})\oplus ({\bf 1},{\bf 1})
\ee
or in a 
\be
{\bf 10_{2/3}} = ({\bf 2},{\bf 2})\oplus ({\bf 1},{\bf 3})\oplus
({\bf 3},{\bf 1})
\ee
With it, KK fermions can be as light as $500~$GeV, and in some cases would have exotic charge assignments. 
The main reason for some KK fermions to be this light has to do with
the need to get the top mass correctly. There are striking signals at
the LHC, for instance from the pair production of charge $5/3$ KK
fermions, a distinct signature for the presence of the extended
custodial symmetry~\cite{ghcust,ghcw}. 

\subsection{Higgs from Fermion Condensation}

Another alternative to dynamically generate the Higgs sector of the  
bulk Randall-Sundrum scenario is the condensation of zero-mode
fermions. Since the localization of fermions near the TeV brane
implies they must have strong couplings to the KK excitations of gauge
bosons, it is possible that the induced four-fermion interaction
is strong enough to result in fermion condensation, thus triggering 
EWSB. Among the SM fermions, the candidate would be the top quark: it
has the strongest localization toward the TeV brane, therefore the
largest coupling to KK gauge bosons, particularly the first excitation
of the KK gluon. The four-fermion interaction would result in a
top-condensation scenario. But we already know that this does not
work if the scale of the underlying interaction is $O(1)~$TeV, as we
expect the KK gluon mass to be. The problem is that the mass
of the condensing fermion should be about $600~$GeV in order for 
the condensation to result in the correct weak scale, $v\simeq
246~$GeV. We can then consider the possibility of 
a fourth generation, one that is highly localized near the TeV brane,
more than the top quark. This results in a effective four-fermion
interaction -mostly mediated by KK gluons-  that is attractive enough 
to trigger the condensation of at least one of the fourth-generation 
quarks. For instance, the up-type fourth-generation quark $U$  
has a four-fermion interaction  given by  
\be
- \frac{g_{01}^L\,g_{01}^R}{M_{KK}^2}\,\left(\bar{U}_L \gamma_\mu t^A U_L\right)\, 
\left(\bar{U}_R \gamma^x\mu t^A U_R\right)~, 
\ee
where $g_{01}^{L,R}$ are the couplings of the left and right-handed
$U$ quarks to the first KK mode of the gluon of mass $M_{KK}$, and $t^A$ are the QCD generators.After Fierz rearrangement, we can re-write this interaction as 
\be
\frac{g_{U}^2}{M_{KK}^2}\,\left\{ \bar{U}^a_L U^a_R \,\,\bar{U}^b_L U^b_R 
- \frac{1}{N_c}\, \bar{U}^a_L U^b_R \,\,\bar{U}^b_L U^a_R \right\}~, 
\label{ffer2terms}
\ee
where $a,b$ are $SU(3)_c$ indices, and  we have defined 
\be
g_{U}^2 \equiv g_{01}^L\,g_{01}^R~.
\ee
The color singlet term in (\ref{ffer2terms}) is attractive, whereas the color octet is 
repulsive, as well as suppressed by $1/N_c$. 
There is a critical value of $g_U^2$ above which there forms a condensate
$
\langle \bar{U}_L\,U_R\rangle$
leading to electroweak symmetry breaking and dynamical masses for the 
condensing fermions. This is 
\be
g_U^2 > \frac{8\pi^2}{N_c}~.
\label{criticality}
\ee
One can also write an effective theory in terms of a scalar doublet 
which becomes dynamical at low energies. So this theory gets a
composite Higgs that is heavy and made of the already mostly-composite
fourth-generation up quarks. The $U$ quark gets a large dynamical
mass. All other zero-mode fermions, including the SM fermions and the 
other fourth-generation zero-modes, get masses through four-fermion
interactions with the $U$ quark. These operators come from 
bulk higher dimensional operators such as 
\be
\int dy\,\sqrt{g}\,\frac{C^{ijk\ell}}{M_P^3}\,\bar{\Psi}^i_L(x,y) \Psi^j_R(x,y) 
\bar{\Psi}^k_R(x,y) \Psi_L^\ell(x,y)
\label{ffermion5d}~, 
\ee
where $C^{ijk\ell}$ are generic coefficients, with $i,j,k,\ell$ standing for  generation 
indices as well as other indices such as isospin, 
and the $\Psi(x,y)$'s can be bulk quarks or leptons. Upon condensation
of the $U$ quarks these result in fermion masses that have the desired
pattern as long as we choose the localization parameters appropriately.

These models have roughly the same $S$ parameter problem as the
generic ones. But to the  tree-level $S$, now we must add also the
loop contributions coming from a heavy Higgs as well as from the
fourth-generation. These are, however, not as large as $S_{\rm tree}$.

The masses of the fourth-generation zero modes could be as small as
$300~$GeV due to mixing with KK modes, and as large as $\sim
600~$GeV. The Higgs is typically rather heavy, in the $(700-900)~$GeV
range. 

The phenomenology of these models is quite different from the three-generation
RS models due to the fact that the fourth generation is the one that
couples strongest to the KK gauge bosons, i.e. to the new physics 
in the s channel~\cite{fgenpheno}. For instance the branching ratio of a KK gauge
boson to $U^{(1)}\bar U^{(1)}$ is likely to be $5-10$ times larger
than that for the $t\bar t$ channel, which is the one always
dominating in three-generation RS models. 
Also the KK gluon tends to be very broad and can only be seen as an
excess in the production of the fourth-generation, which is dominated
by QCD. In general, all KK gauge bosons will be considerably broader,
although electroweak KK gauge bosons might have more manageable
widths. 

\section{Conclusions}
Model building in a slice of AdS$_5$ extends the Randall-Sundrum solution to the 
hierarchy problem, and opens up the possibility of addressing dynamically the origin of 
EWSB and fermion masses, among other things. Through the AdS/CFT correspondence we can see 
that these weakly coupled 5D theories are dual to strongly coupled 4D gauge theories. 
Thus, building bulk RS theories of EWSB corresponds, through holography, to certain strongly 
coupled electroweak sectors. Although the type of strongly coupled theory is not completely generic, 
this procedure gives us access to a large variety of strongly coupled theories of the electroweak 
sector. 

In addition to solving the hierarchy problem, RS bulk models provide a framework to 
understand the hierarchy of fermion masses. This implies the presence of tree-level flavor 
violation with KK gauge bosons. Rigorous compatibility with low energy data may require some level 
of flavor symmetry in the bulk. However, this is not surprising since fermion localization already 
signaled some amount of flavor breaking in the UV. The central point is that the metric in AdS$_5$ provides
the necessary large scale separation between the light and the heavy fermions. 

Finally, there are several alternatives for the Higgs sector. 
Higgsless models are viable, although require some measure of fine-tuning to cancel
contributions to $Z\to b_L\bar b_L$.
Gauge-Higgs unification models are very promising and in best agreement with electroweak precision
constraints. They provide a dynamical origin for the localization of the Higgs near the TeV brane, 
and in the Holographic dual correspond to a composite pNGB Higgs. 
Finally, another alternative to localize a composite Higgs near the TeV brane, is the condensation 
of  fourth-generation  quarks via the attractive four-fermion interaction mediated mostly by the 
KK gluon. This is a distinct possibility, a realization of the fourth-generation condensation 
proposed long ago by Bardeen, Hill and Lindner~\cite{bhl}. These three Higgs sector possibilities have very different
phenomenology and should be distinguishable at the LHC.

\section*{Acknowledgements} The author acknowledges the support of the State of São Paulo
Research Foundation (FAPESP), and the Brazilian  National Counsel
for Technological and Scientific Development (CNPq).

\end{document}